# Control Efficacy on COVID-19

Duanbing Chen[*], Tao Zhou[†]
Big Data Research Center, University of Electronic Science and Technology of China, Chengdu, People's Republic of China

**Abstract**. We proposed a Monte-Carlo method to estimate temporal reproduction number without complete information about symptom onsets of all cases. Province-level analysis demonstrated the huge success of Chinese control measures on COVID-19, that is, provinces' reproduction numbers quickly decrease to <1 by just one week after taking actions.

Emerged from Wuhan City, the novel coronavirus diseases rapidly expanded since December 2019, and slowed down recently in mainland China. Early analyses indicated that COVID-19 has middle-to-high transmissibility, with preliminary estimates of basic reproduction number $R_0$ lying in the range [2.0, 4.0], e.g., 1.4-3.9 (1), 2.47-2.86 (2) and 2.8-3.9 (3). After a period of stealthy spread, on Jan 20, 2020, COVID-19 pneumonia was identified as a B-type infectious disease in China, and the control measures were set according to the standard of A-type infectious disease. Roughly speaking, Jan 21 can be considered as the starting date of control, on which every province in China took COVID-19 spread as an emergency event and launched strong control measures. These control measures have achieved remarkable success, with daily number of confirmed cases quickly

---

[*] Dr. Chen is a full professor in the School of Computer Science and Engineering at the University of Electronic of Science and Technology of China. His main research interests are structure and dynamics of complex networks.
[†] Corresponding author (zhutou@ustc.edu).

decreasing after a short expansion lasting about two weeks from Jan 21.

The most intuitive metric to quantify the control efficacy is the effective reproduction number $R_t$, which is defined as the mean number of secondary cases infected by a case with symptom onset at day $t$. Here we consider a slightly different one, temporal reproduction number, to include the period-dependent metric $R_{[t_1,t_2]}$ ($t_1 \leq t_2$) that is defined as the mean number of secondary cases infected by a case with symptoms onset during the time period $[t_1,t_2]$ (4). Accordingly, $R_t$ is a special case of $R_{[t_1,t_2]}$ when $t_1 = t_2 = t$.

If complete information about who infected whom is known, the reproduction number can be determined by simply counting secondary cases. However, tracing information is usually incomplete or not timely available, and thus statistical approaches are required. Willinga and Teunis (5) proposed a likelihood-based method to estimate $R_t$ from the epidemic curve and the distribution of generation intervals, which works only for period from which all secondary cases would have been detected, thus resulting in a time lag about 19 days for COVID-19 (95th percentile of the distribution of generation intervals) (1). By accounting for yet unobserved secondary cases via Bayesian inference, Cauchemez et al. (6) extended Wallinga-Teunis method to provide real-time estimates of $R_t$.

Here we consider an even-worse condition about data accessibility, where not only the complete tracing records, but also the full epidemic curves are unknown. That is a usual situation in the early stage of an epidemic, for example, the number of confirmed cases of COVID-19 for each province in mainland China is made public every day, while the symptom onset of each case is not reported by Chinese CDC. We develop a Monte-Carlo method to infer the epidemic curve from a small number of recorded symptom onsets collected from scattered news reports (in total we have collected 3650 records with precise symptom onset time). Combining it with the methods proposed in (5) and (6), we can estimate temporal reproduction number and thus evaluate the efficacy of control measures. Technical details are given in (Appendix).

In addition to assumptions in (6), our method depends on another assumption that the distribution of the time intervals between symptom onsets and confirmations for each province, $p(t_\Delta)$, is close to the synthesized one by scattered records. Based on the six provinces with the most records of symptom onsets, we have checked that the individual distributions are close to each other and can be well resembled by the synthesized distribution, which follows a translational Weibull distribution (Appendix). The province-level results (Table) demonstrate the impressive achievement of control measures, namely $R_t$ for the majority of provinces decreased to <1 within one week from the starting date of control. Even for Hubei, the epidemic was under control ($R_t < 1$) just in two weeks. In addition, the average temporal reproduction number $R_{[Feb\ 15,\ Feb\ 21]}$ over all provinces already decayed to 0.18, a very small value corresponding to a dying phase of the epidemic. This method can be further improved by considering importations (7,8) and using Markov-Chain Monte-Carlo algorithm based on independent transmission assumption (9,10). Discussion on limitations are presented in (Appendix).

**Table**. Results for all provinces in mainland China except Tibet and Qinghai, where the confirmed cases are too few to do statistics. For each province, we show: (i) the number of cumulated confirmed cases by Feb 22, 2020; (2) the date $t^*$ when $R_t$ below 1; and (iii) the temporal reproduction number during the last week [Feb 15, Feb 21]. The results are averaged over 10000 independent runs. Detailed results for all provinces are shown in (Appendix).

| Province | Number of cumulated confirmed cases | Date $t^*$ when $R_t$ below 1 | Temporal reproduction number of the last week |
|---|---|---|---|
| Fujian | 298 | 2020/1/23 | 0.1365 |
| Liaoning | 121 | 2020/1/23 | 0.0053 |
| Yunnan | 174 | 2020/1/23 | 0.2039 |
| Shanghai | 335 | 2020/1/24 | 0.1967 |
| Zhejiang | 1205 | 2020/1/24 | 0.2895 |
| Chongqing | 573 | 2020/1/24 | 0.2463 |
| Beijing | 399 | 2020/1/25 | 0.2493 |
| Gansu | 91 | 2020/1/25 | 0.0000 |
| Guangdong | 1342 | 2020/1/25 | 0.1088 |
| Guangxi | 249 | 2020/1/25 | 0.3232 |
| Hunan | 1016 | 2020/1/25 | 0.1321 |
| Shaanxi | 245 | 2020/1/25 | 0.3002 |
| Sichuan | 526 | 2020/1/25 | 0.1757 |
| Henan | 1271 | 2020/1/26 | 0.0848 |
| Nei Monggol | 75 | 2020/1/26 | 0.3176 |
| Ningxia | 71 | 2020/1/26 | 0.0146 |
| Shanxi | 132 | 2020/1/26 | 0.2780 |
| Shandong | 754 | 2020/1/27 | 0.4977 |
| Anhui | 989 | 2020/1/27 | 0.0820 |
| Hainan | 168 | 2020/1/27 | 0.3487 |
| Jiangsu | 631 | 2020/1/27 | 0.0901 |
| Jiangxi | 934 | 2020/1/27 | 0.0556 |
| Tianjin | 135 | 2020/1/27 | 0.4241 |
| Hebei | 311 | 2020/1/28 | 0.1736 |
| Jilin | 91 | 2020/1/28 | 0.1651 |
| Guizhou | 146 | 2020/1/29 | 0.0156 |
| Heilongjiang | 480 | 2020/1/29 | 0.1307 |
| Xinjiang | 76 | 2020/1/30 | 0.1320 |
| Hubei | 64287 | 2020/2/2 | 0.0491 |

Via estimating the province-level temporal reproduction number, this Letter has demonstrated the remarkable efficacy of Chinese control measures on COVID-19, which is obviously resulted from the ambitious and aggressive government-led actions, as well as the high efficiency of the hierarchical structure of Chinese leadership. However, what we would like to emphasize lastly is that advanced information techniques are widely employed in China to trace the epidemic spreading. For example, in many cities, QR codes are posted in buses, subway stations, taxies, supermarkets, bazaars, hotels, restaurants, office buildings, and so on, and people are asked to scan the codes (check-in with mobile phones) before entering (Appendix). Therefore, after laboratory confirmation of any case, the administrators know immediately and exactly the persons who have possible contacts with this case. This is a perfect tool in the epidemiological perspective to efficiently and effectively block the spread through communities. We hope other countries suffering COVID-19 epidemics would learn from Chinese practices.

## Acknowledgments

We thank Yan Wang, Wei Bai and Min Wang for data collection, Qin Gu for developing the check-in system via scanning the QR codes and sharing some representative QR codes with us, and Quan-Hui Liu for helpful discussion. This work is partially supported by the National Natural Science Foundation of China under grant numbers 61673085, 11975071 and 61433014.

# Appendix of "Control Efficacy on COVID-19"

## Appendix Section 1. Methods

First of all, using the collected records from scattered new reports, we can obtain the synthesized distribution $p(t_\Delta)$, where $t_\Delta$ denote the time interval between symptom onset and laboratory confirmation. Then, given a case $i$ confirmed at day $t^{(i)}$, we can sample a time interval $t_\Delta^{(i)}$ according to $p(t_\Delta)$ and set $i$'s symptom onset time as $t_i = t^{(i)} - t_\Delta^{(i)}$. For each province, in each run, we apply such Monte-Carlo sampling method to approximately estimate all confirmed cases' symptom onsets. In this Letter, we implement $10^4$ independent runs to obtain the mean values and confidence intervals.

According to the empirical observations (1), the distribution of generation intervals, $q(t_g)$, is approximated by a Gamma distribution

$$q(t_g) = \frac{\beta^\alpha}{\Gamma(\alpha)} t_g^{\alpha-1} e^{-\beta t_g} \quad (t_g > 0),$$

where the shape parameter $\alpha = 4.866$, and the inverse scale parameter $\beta = 0.649$. Given two cases $i$ and $j$ with $t_i > t_j$, the likelihood that case $i$ is infected by case $j$ is

$$\rho_{ij} = \frac{q(t_i - t_j)}{\sum_{k, t_i > t_k} q(t_i - t_k)}.$$

Wallinga and Teunis (5) suggested that the expected number of secondary cases infected by case $j$ can be estimated by the sum of likelihoods, as

$$R_j = \sum_{i, t_i > t_j} \rho_{ij}.$$

The effective reproduction number can thus be estimated as

$$R_t = \frac{1}{|C_t|} \sum_{j \in C_t} R_j,$$

where $C_t$ is the set of cases with symptom onsets at day $t$. Obviously, $R_t = R_j$ if $j \in C_t$ since in the Wallinga-Teunis method, cases with the same symptom onset time have the same expected number of secondary cases. Analogously, the temporal effective number can be estimated as

$$R_{[t_1, t_2]} = \frac{1}{|C_{[t_1, t_2]}|} \sum_{j \in C_{[t_1, t_2]}} R_j,$$

where $C_{[t_1, t_2]}$ is the set of cases with symptom onsets in the range $[t_1, t_2]$.

We further consider the task to calculate the effective reproduction number $R_t$ given the last known onset time being day $T$. Obviously, only if $T > t$, this task is possible. If $T \geq t + t_g^{\max}$ with $t_g^{\max}$ denoting the maximum generation interval, we can directly apply the Wallinga-Teunis method. However, if $t < T < t + t_g^{\max}$, we need to introduce an additional step with Bayesian inference (6). Assuming the mean number of secondary cases infected by a case with symptom onset at day $t$ can be decomposed by two parts as

$$R_t = R_t^-(T) + R_t^+(T),$$

where $R_t^-(T)$ and $R_t^+(T)$ are the mean numbers of secondary cases with symptom onsets before or at $T$ and after $T$, respectively. The value of $R_t^-(T)$ can be directed estimated by using the Wallinga-Teunis method, and thus we can

infer the effective reproduction number as

$$R_t = \frac{R_t^-(T)}{\sum_{t_g=1}^{T-t} q(t_g)}.$$

The temporal reproduction number for a given time period can be obtained in a similar way.

## Appendix Section 2. Synthesized Distribution $p(t_\Delta)$

Appendix Figure 1 compares the distributions of time intervals between symptom onsets and confirmations for the six provinces with most known records, indicating that the synthesized distribution can well resemble the province-level distributions.

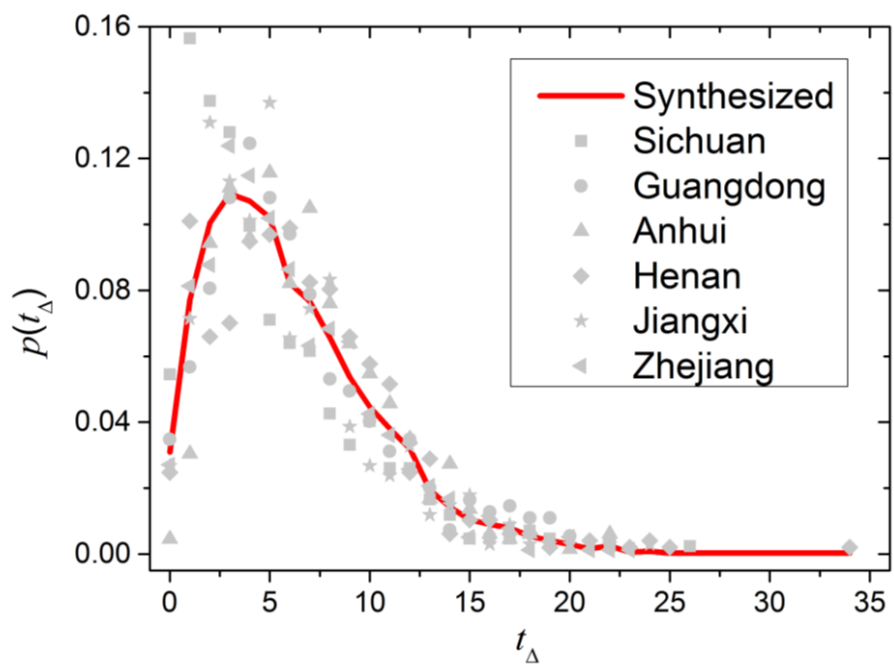

**Appendix Figure 1**. Comparison between the synthesized distribution of time intervals between symptom onsets and confirmations (red solid line) and the individual distributions of Sichuan, Guangdong, Anhui, Henan, Jiangxi and Zhejiang (gray data points).

As shown in Appendix Figure 2, the synthesized distribution $p(t_\Delta)$ can be well fitted by a translational Weibull distribution

$$p(t_\Delta) = \frac{\alpha}{\beta}\left(\frac{t_\Delta+\gamma}{\beta}\right)^{\alpha-1} e^{-\left(\frac{t_\Delta+\gamma}{\beta}\right)^\alpha},$$

where the shape parameter $\alpha=1.48$, the scale parameter $\beta=7.03$, and the translational parameter $\gamma=0.10$. We introduce the translational parameter because some cases are confirmed immediately so $p(0)>0$, while the original Weibull distribution gives $p(0)=0$ for any shape parameter and scale parameter.

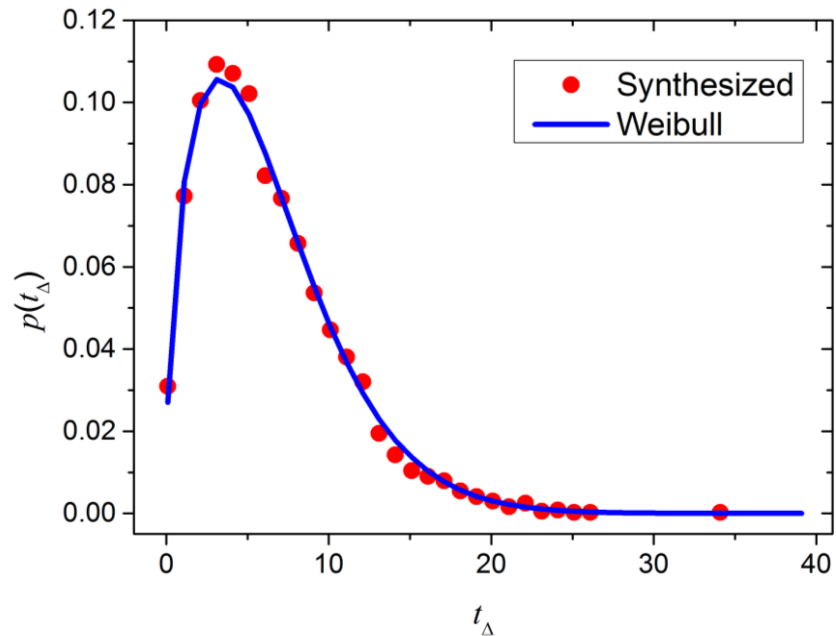

**Appendix Figure 2.** Comparison between the synthesized distribution of time intervals between symptom onsets and confirmations (red circles) and the fitting curve (blue curve) that obeys a translational Weibull distribution.

## Appendix Section 3. Detailed Results for All Provinces

We have collected the information about daily number of confirmed cases for all provinces in mainland China from Jan 11, 2020 to Feb 22, 2020. The number of cumulated confirmed cases is 76,936. For a very small fraction (4.74%) of these confirmed cases, we have found their symptom onset times by hand via scattered new reports. The confirmed cases for Tibet and Qinghai are only 1 and 15, so we do not analyze these two provinces. Appendix Figure 3 reports the estimated effective production number for each province from Jan 10, 2020 to Feb 21, 2020 by using the present method.

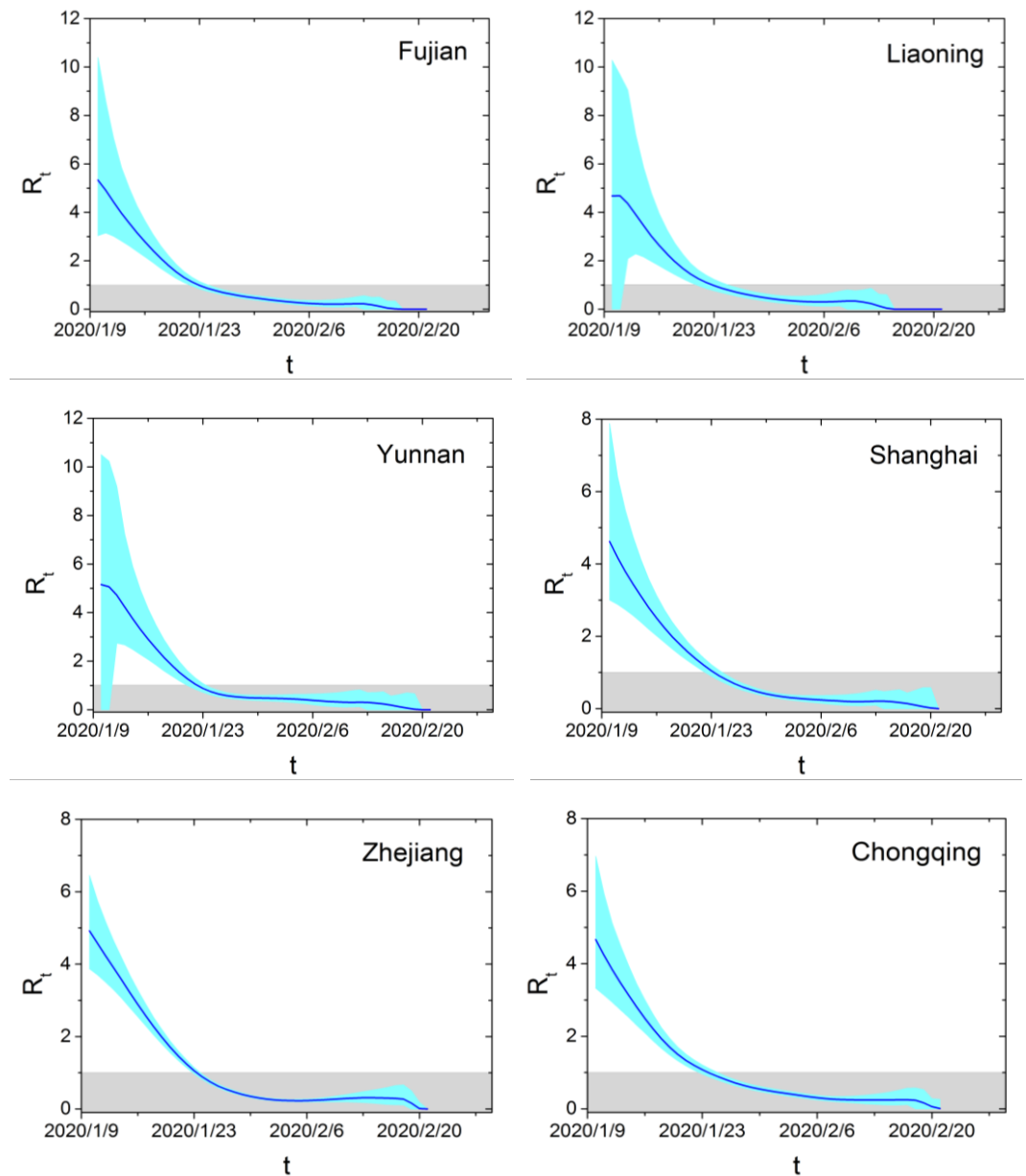

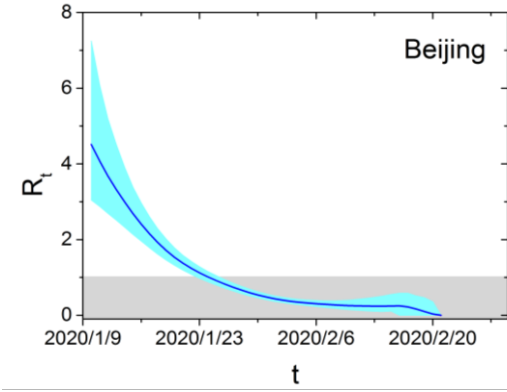
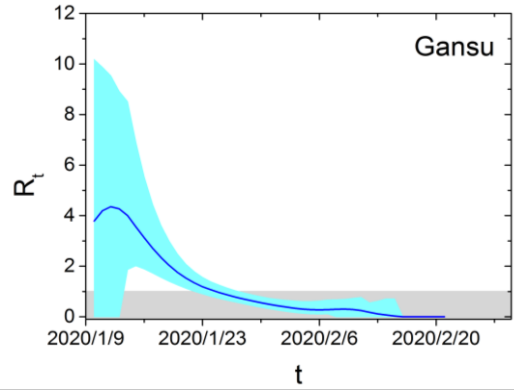
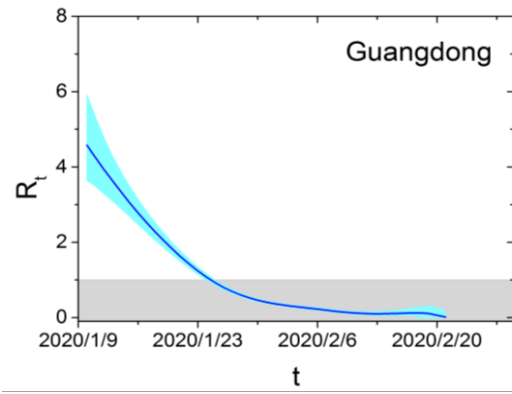
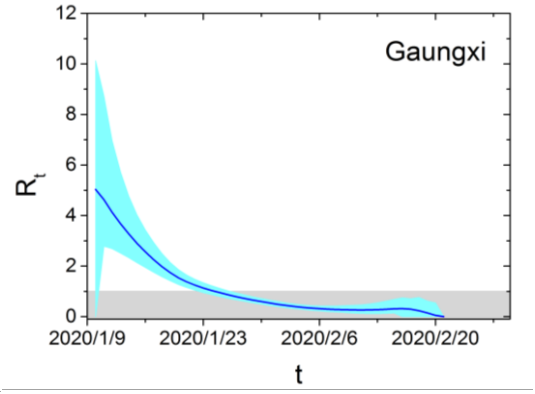
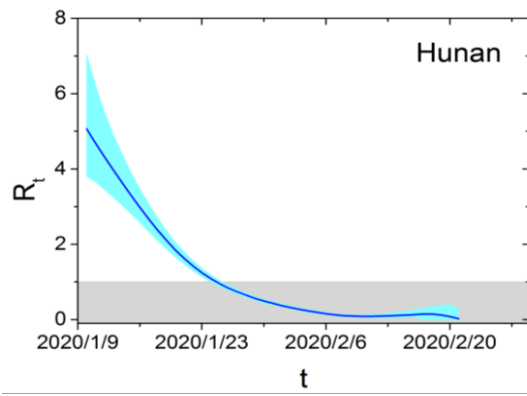
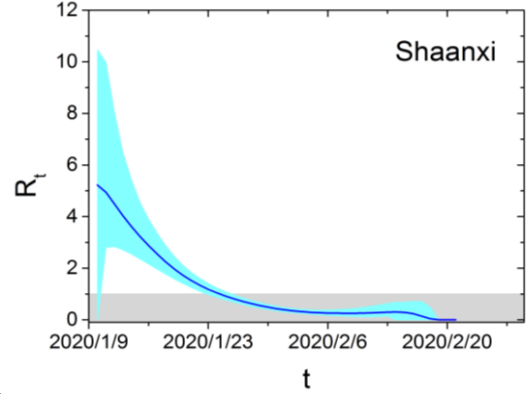
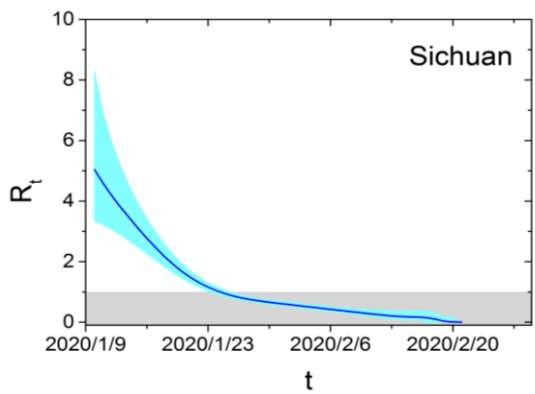
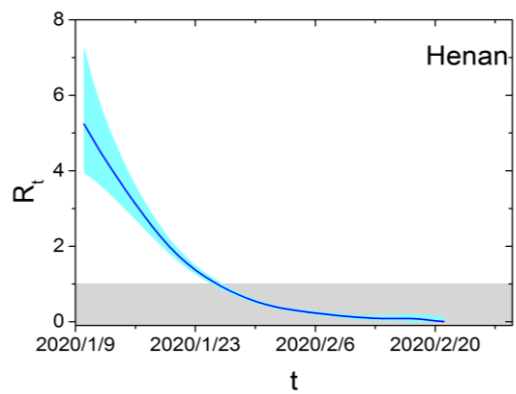

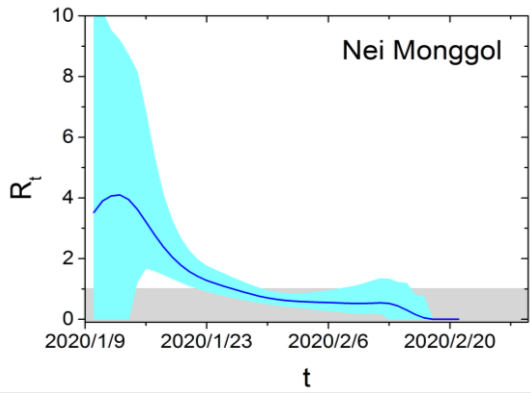
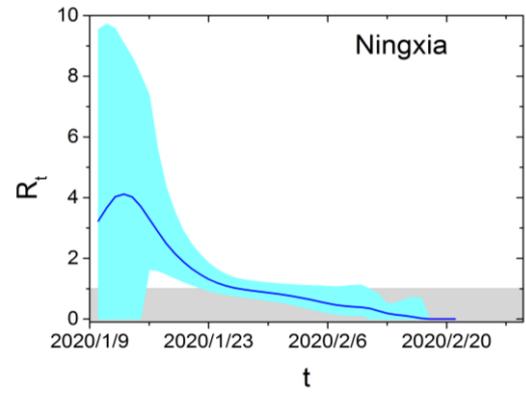
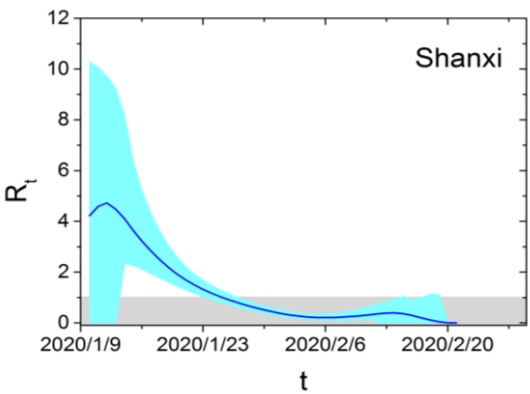
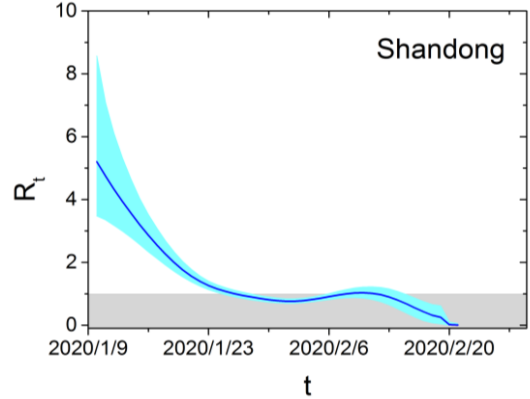
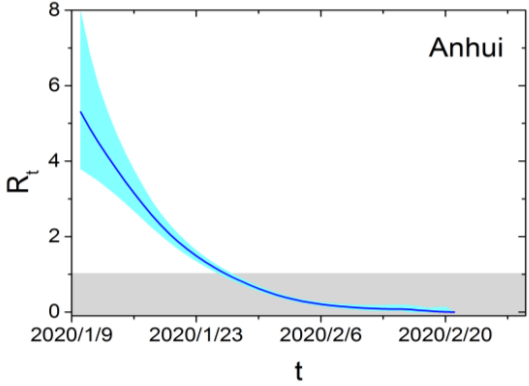
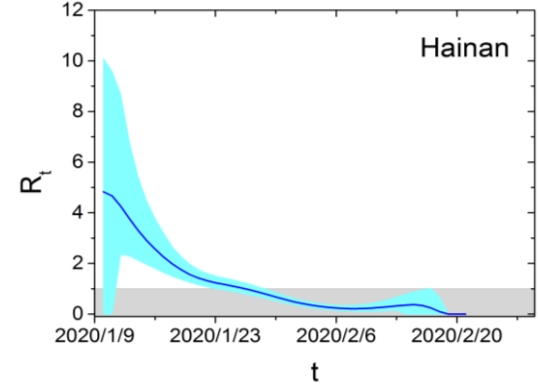
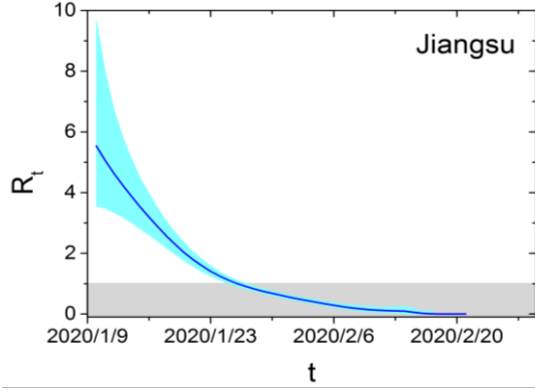
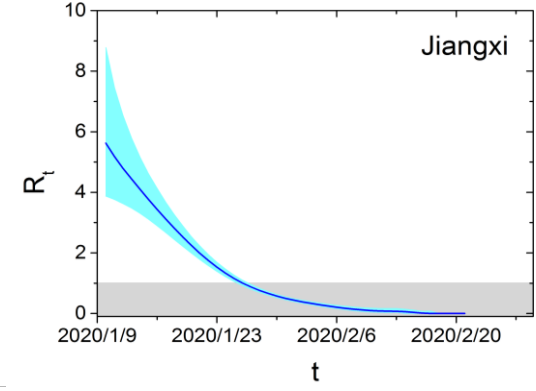

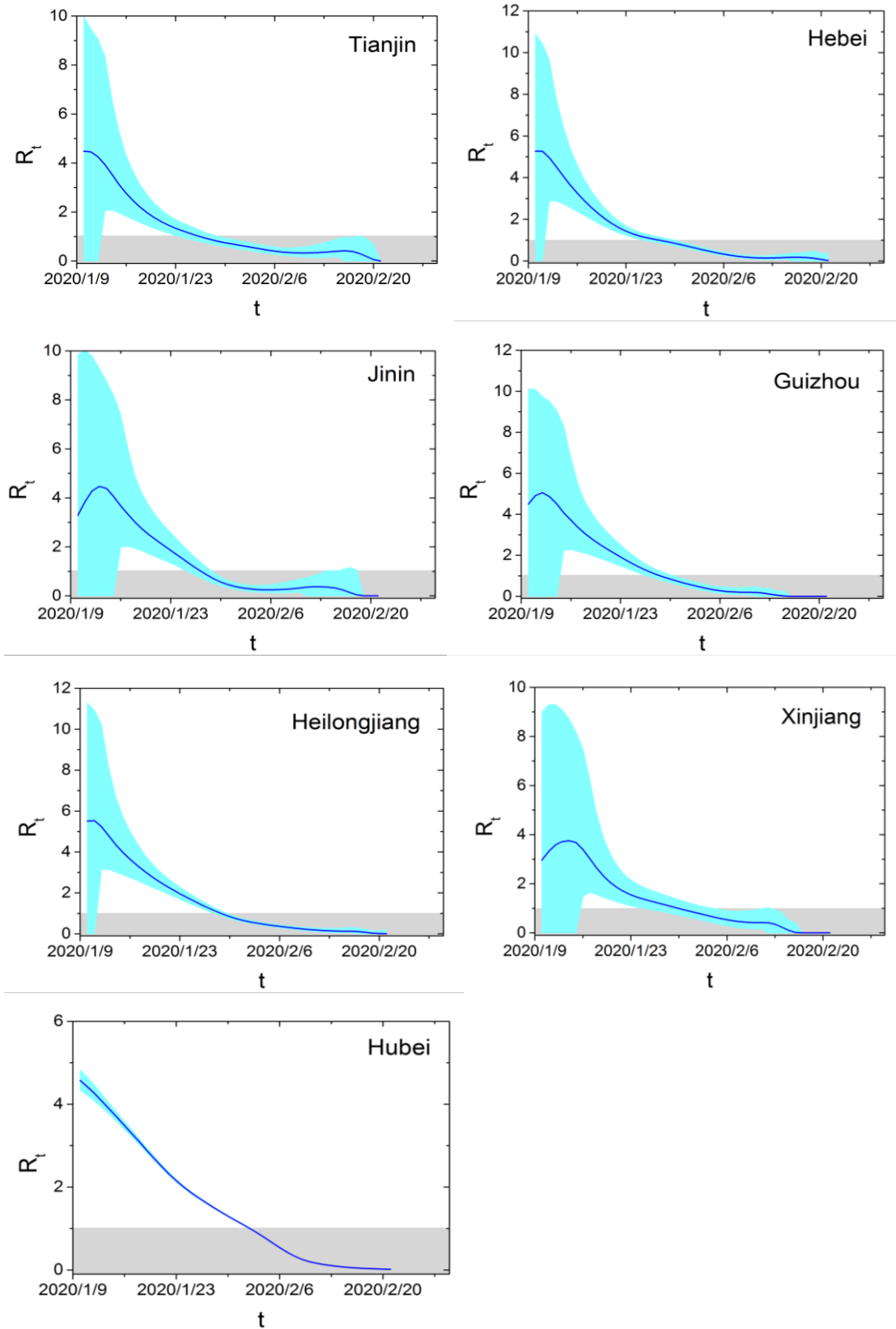

**Appendix Figure 3**. Effective reproduction numbers for all provinces in mainland China from Jan 10, 2020 to Feb 21, 2020. The results are averaged over 10000 independent runs, and the cyan areas denote the 95% confidence intervals. In each run, the Monte-Carlo sampling method is applied to infer the symptom onset times. The gray shadows emphasize the situations where the epidemic is under control ($R_t<1$).

# Appendix Section 4. Limitations

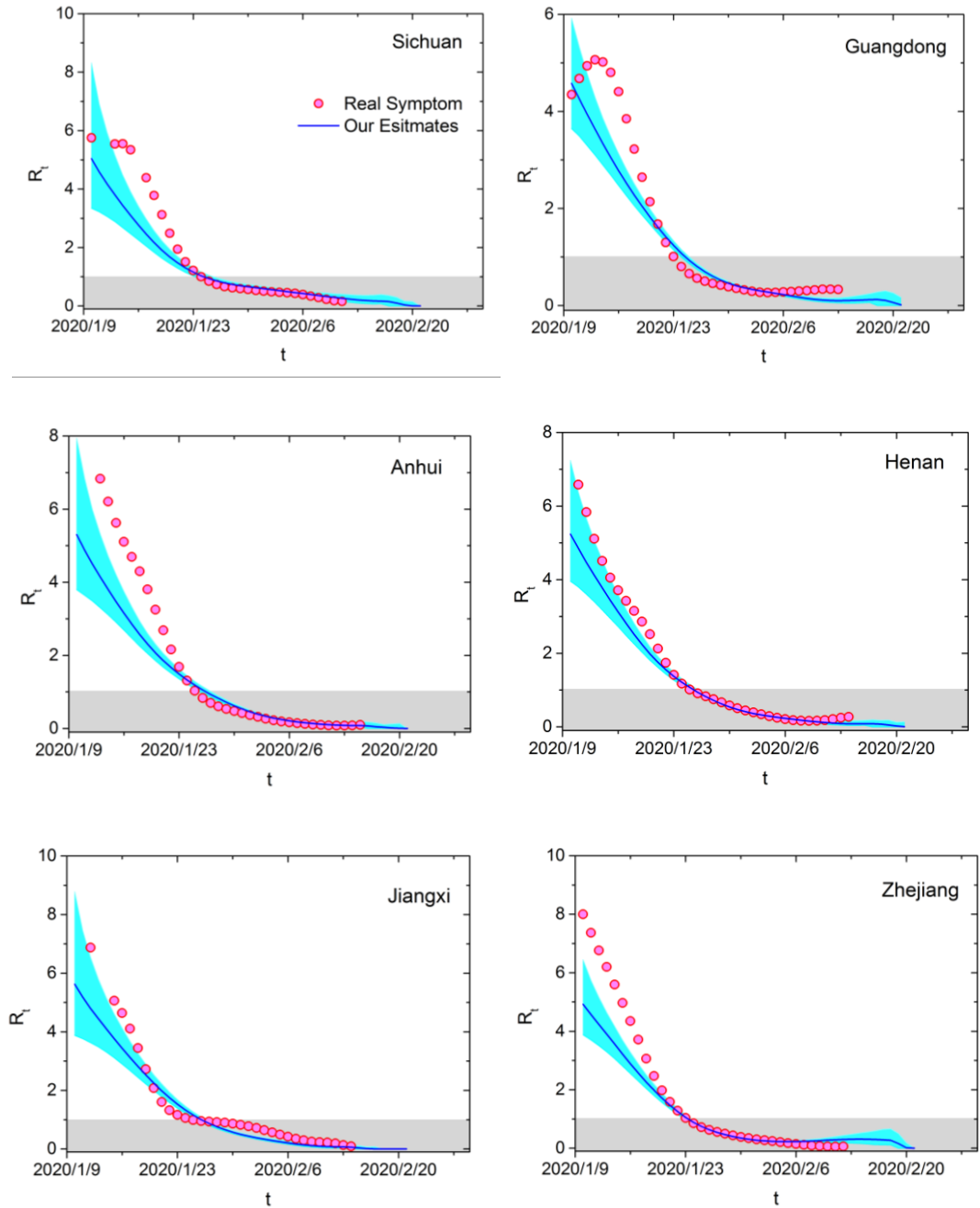

**Appendix Figure 4**. Comparison between the estimates of effective reproduction numbers by the true and inferred records of symptom onsets. The solid blue curves and cyan areas respectively denote the average values and 95% confidence intervals obtained by 10000 independent runs according to the inferred data. The red circles represent the results obtained by the true records. The gray shadows emphasize the situations where the epidemic is under control ($R_t<1$). The six plots are results for Sichuan, Guangdong, Anhui, Henan, Jiangxi and Zhejiang.

Though the synthesized distribution $p(t_\Delta)$ can well resemble individual distributions, using inferred data may still bring bias because the distribution $p(t_\Delta)$ is not stable, usually with smaller and smaller mean and standard deviation in the progress of an epidemic (8). Appendix Figure 4 compares the estimates of effective reproduction numbers by the true and inferred records of symptom onsets for the six provinces with most known records. At the very beginning, the estimates from inferred data are smaller than the ones from true records, but they are getting closer and closer later and show almost the same $t^*$.

Indeed, we still overestimate the reproduction number in the early stage, because a large fraction of cases (expect Hubei) are importations (7,8). Fortunately, the present method shows accordance with the one accounting for importations in the later part, for example, $R_t$ of the three example provinces (Guangdong, Hunan and Shandong) approach zero and then continuously decrease at Jan 23, Jan 26 and Jan 30 by the method in (8) and at Jan 25, Jan 25, Jan 27 by the present method.

## Appendix Section 5. Example of QR Codes

Appendix Figure 5 illustrates an example the QR codes used to trace the trajectories of people. These codes are posted mainly in the public transport means and places with crowds. The example shown here was posted in a bus in Chengdu City of Sichuan Province, and people are required to scan the code before getting on the bus. Therefore, if a confirmed or suspected case has taken this bus, we can immediately find out people who has also taken this bus in the same time period. QR codes posted in other places play similar roles.

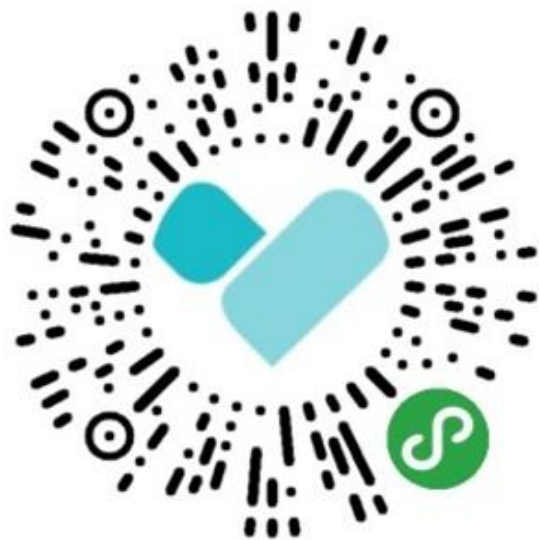

**Appendix Figure 5**. Illustration of an example of the QR codes to trace the epidemic in mainland China. This is the one posted in a public bus in Chengdu City. In the bottom, a Chinese character followed by A11345 is the plate number of this bus, and the character is the abbreviation of Sichuan Province.